\documentclass[preprint,3p]{elsarticle}

\usepackage{enumerate}
\usepackage{amssymb}
\usepackage{amsthm}
\usepackage[mathlines]{lineno}
\usepackage{graphicx}
\usepackage{units}
\usepackage{url}
\usepackage{amsmath}
\usepackage{amsfonts}
\usepackage{bm}
\usepackage{textcomp}
\usepackage{subfigure}
\usepackage{multicol}
\usepackage{multirow}
\usepackage{verbatim}
\usepackage{rotating}
\usepackage[colorlinks,linkcolor=red,citecolor=red]{hyperref}
\usepackage{float}
\usepackage{epsfig}
\usepackage{dcolumn}
\usepackage{bm}
\usepackage{color}
\usepackage{pstricks}
\usepackage{pst-node}
\usepackage{times}
\usepackage{indentfirst}
\usepackage{overpic}
\usepackage[english]{babel}
\addto{\captionsenglish}{%
  
  \renewcommand{\tablename}{TABLE}
}
\journal{Physics Letters B}

\newcommand{\tto}{\rightarrow}

\newcommand{\ee}{e^+e^-}

\newcommand{\phione}{\phi_{1}}
\newcommand{\phitwo}{\phi_{2}}

\begin{document}
\begin{frontmatter}
\title{{\bf \boldmath Measurement of cross sections of the interactions $e^+e^-\rightarrow \phi\phi\omega$ and $e^+e^-\rightarrow \phi\phi\phi$ at center-of-mass energies from 4.008 to 4.600\,GeV}}
\author{
M.~Ablikim$^{1}$, M.~N.~Achasov$^{9,e}$, S. ~Ahmed$^{14}$, X.~C.~Ai$^{1}$, O.~Albayrak$^{5}$, M.~Albrecht$^{4}$, D.~J.~Ambrose$^{44}$, A.~Amoroso$^{49A,49C}$, F.~F.~An$^{1}$, Q.~An$^{46,a}$
% ...
\\
\vspace{0.2cm}
(BESIII Collaboration)\\
\vspace{0.2cm} {\it
$^{1}$ Institute of High Energy Physics, Beijing 100049, People's Republic of China\\
% ...
$^{4}$ Bochum Ruhr-University, D-44780 Bochum, Germany\\
$^{5}$ Carnegie Mellon University, Pittsburgh, Pennsylvania 15213, USA\\
% ...
$^{9}$ G.I. Budker Institute of Nuclear Physics SB RAS (BINP), Novosibirsk 630090, Russia\\
% ...
$^{14}$ Helmholtz Institute Mainz, Johann-Joachim-Becher-Weg 45, D-55099 Mainz, Germany\\
% ...
$^{44}$ University of Rochester, Rochester, New York 14627, USA\\
% ...
$^{46}$ University of Science and Technology of China, Hefei 230026, People's Republic of China\\
% ...
$^{49}$ (A)University of Turin, I-10125, Turin, Italy; (B)University of Eastern Piedmont, I-15121, Alessandria, Italy; (C)INFN, I-10125, Turin, Italy\\
% ...
$^{a}$ Also at State Key Laboratory of Particle Detection and Electronics, Beijing 100049, Hefei 230026, People's Republic of China\\
% ...
$^{e}$ Also at the Novosibirsk State University, Novosibirsk, 630090, Russia\\
% ...
}
}

\begin{abstract}
Using data samples collected with the BESIII detector at the BEPCII collider at six center-of-mass energies between 4.008 and 4.600\,GeV, we observe the processes $e^+e^-\rightarrow \phi\phi\omega$ and $e^+e^-\rightarrow \phi\phi\phi$. The Born cross sections are measured and the ratio of the cross sections $\sigma(e^+e^-\rightarrow \phi\phi\omega)/\sigma(e^+e^-\rightarrow \phi\phi\phi)$ is estimated to be $1.75\pm0.22\pm0.19$ averaged over six energy points, where the first uncertainty is statistical and the second is systematic. The results represent first measurements of these interactions.
\\

\end{abstract}
\end{frontmatter}

\begin{multicols}{2}
\section{Introduction}
The experimental understanding of hadron production in electron-positron annihilation has been achieved with the measurement of the total inclusive hadronic cross sections, the so-called R measurement~\cite{Bes3-R}, and the exclusive measurement of final states involving pions, kaons and other light hadrons at various center-of-mass (c.m.) energies~\cite{exh-Babar,exh-BESIII}. The tools for describing the $e^+e^-$ annihilation to hadrons process generally include the use of the KKMC generator~\cite{KKMC}, which includes initial and final state radiation, and the Pythia~\cite{Pythia} program based on the Lund String model or Parton Shower model that hadronizes the final-state quarks. The KKMC-Pythia combination is not expected to correctly describe the processes with more than two vector mesons in the final state, as they correspond to higher order Quantum Chromodynamics (QCD) processes and are generally associated with multiple gluons. The experimental results provide more constraints on the higher-order QCD calculation.

The BaBar and Belle collaborations reported the observation of significant double charmonium production $\ee\tto J/\psi c\bar{c}$ and found the ratio $\sigma(\ee\tto J/\psi c\bar{c}) / \sigma(\ee\tto J/\psi X)$ to be $\sim0.6$~\cite{Belle-jpsiccbar}, which indicates that a surprisingly large fraction of $\ee\tto J/\psi X$ events are produced by the $\ee\tto J/\psi c\bar{c}$ process.
This experimental result has stimulated much theoretical interest. Various theoretical approaches, such as NRQCD factorization~\cite{NRQCD} and the light cone method~\cite{NLO}, have been proposed to make corrections to the low ratio predicted by the non-relativistic calculation, which predicts a much lower value for the cross section~\cite{pred-Jpsiccbar}. The validity of the theoretical investigations can be tested over a wide kinematical range with double or triple quarkonia ($s\bar{s}$, $c\bar{c}$, $b\bar{b}$) produced in $e^+e^-$ annihilations. In particular, strangeonia $s\bar{s}$ are located in the region of transition between perturbative QCD and non-perturbative QCD. The $e^+e^-$ annihilation to multiple $s\bar{s}$ states may provide an important experimental opportunity in the low-energy region.

In this paper, we report on the first measurement of the Born cross sections of $e^+e^-\rightarrow \phi\phi\omega$ and $e^+e^-\rightarrow \phi\phi\phi$ processes at c.m.\ energies $E_{\rm{cm}}=4.008, 4.226, 4.258, 4.358, 4.416$ and $4.600$\,GeV~\cite{Ecm}. The data samples were collected by the BESIII detector at the BEPCII collider~\cite{BEPCII}.

\begin{figure*}
\begin{center}
\includegraphics[width=0.45\linewidth]{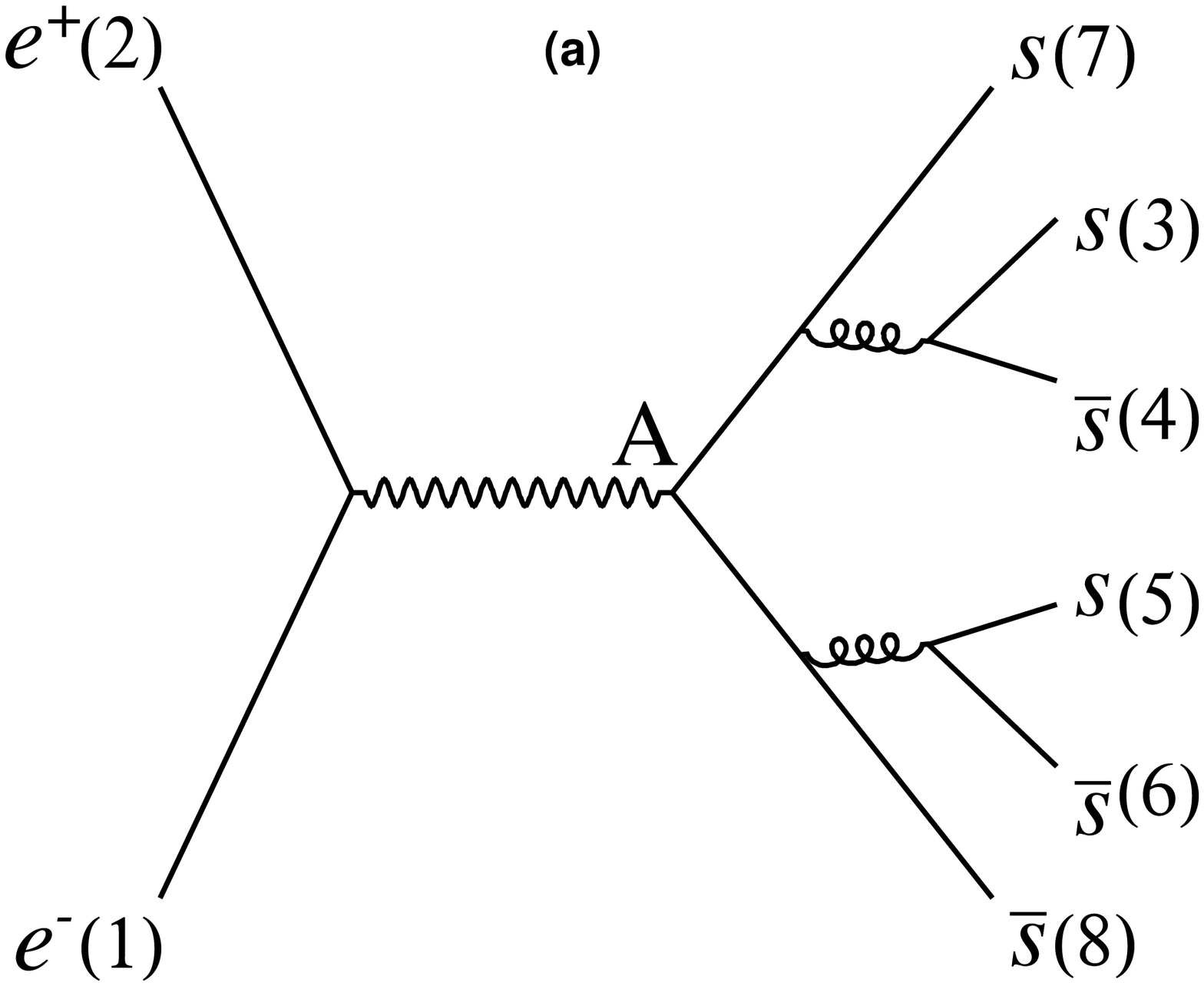}
\includegraphics[width=0.45\linewidth]{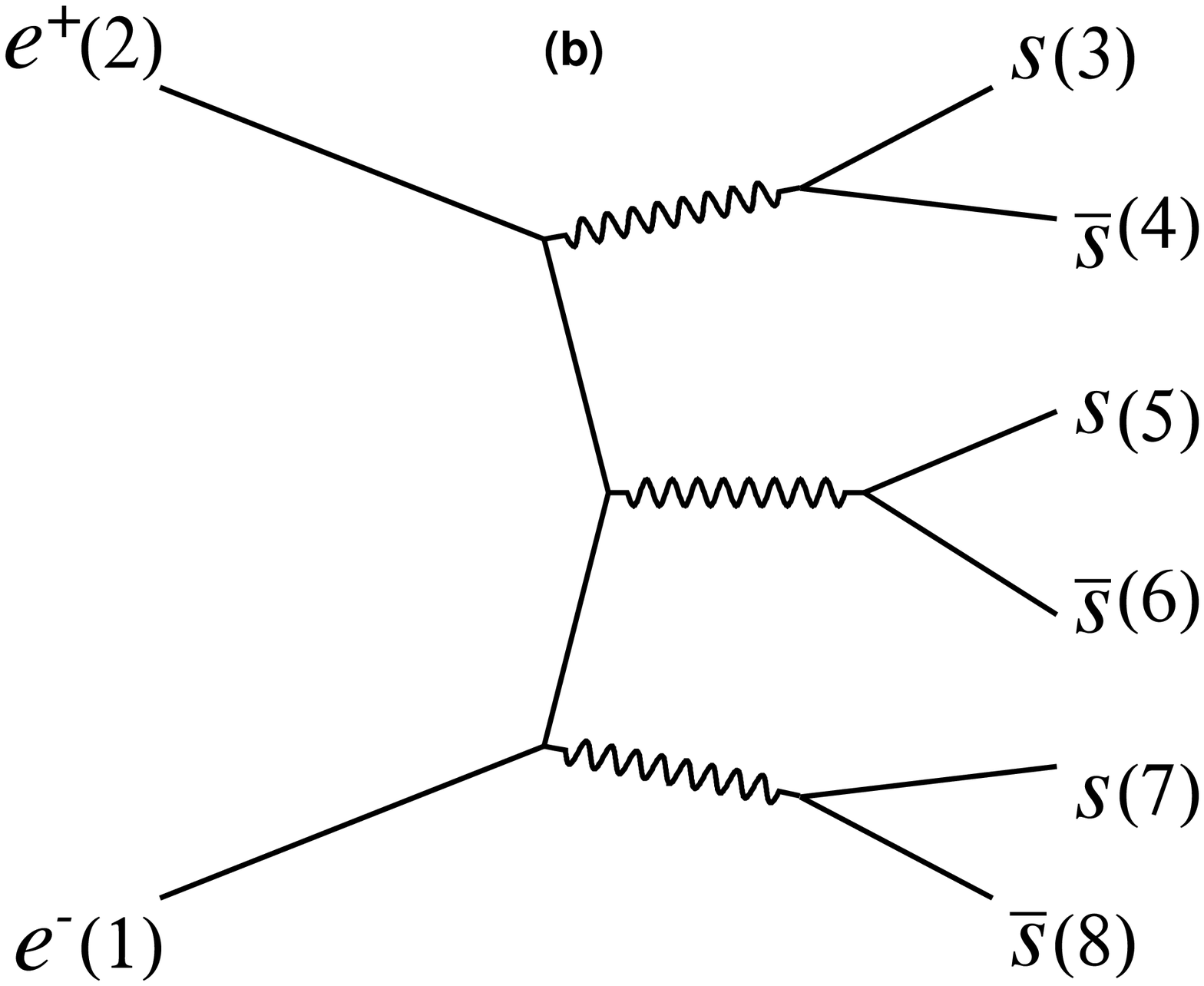}
\caption{Feynman graphs for (a) $e^+e^-\rightarrow \gamma^*gg \rightarrow 3(s\bar{s})$. (b) $e^+e^-\rightarrow 3\gamma^* \rightarrow 3(s\bar{s})$.}
\label{fig:FMD-phiphiX}
\end{center}
\end{figure*}

Additionally, we also measure the ratio $\sigma(e^+e^-\rightarrow \phi\phi\omega)/\sigma(e^+e^-\rightarrow \phi\phi\phi)$, where many of the systematic uncertainties are canceled. The mixing angle of $\omega$ and $\phi$ is expected to be small and its effect on the ratio can be neglected. In the $e^+e^-$ annihilation process, without considering the intermediate resonance, the final $\phi\phi\phi$ states would be generated via one virtual photon and two gluons or three virtual photons, as illustrated in FIG.~\ref{fig:FMD-phiphiX}. The production via two virtual photons and one gluon is forbidden, because the gluon carries color while the final state is color neutral. By replacing $s(7)\bar{s}(8)$ with $(u\bar{u}+d\bar{d})/\sqrt{2}$ in Fig.~\ref{fig:FMD-phiphiX}(a), we obtain the ratio $\frac{\sigma(e^+e^-\rightarrow \gamma*gg \rightarrow 2(s\bar{s})+(u\bar{u}+d\bar{d})/\sqrt{2})} {\sigma(e^+e^-\rightarrow \gamma*gg \rightarrow 3(s\bar{s}))}\sim\frac{(\frac{4}{9}+\frac{1}{9})/2}{\frac{1}{9}}=2.5$, because the vertex ``A" is proportional to the charge squared of the quarks. If, on the other hand, $(u\bar{u}+d\bar{d})/\sqrt{2}$ is substituted for $s(3)\bar{s}(4)$ or $s(5)\bar{s}(6)$, the ratio would be about 1 since the strong interaction vertex only relies on the mass of the quarks. Considering the above two cases in
Fig.~\ref{fig:FMD-phiphiX}(a) and neglecting the small contribution from Fig.~\ref{fig:FMD-phiphiX}(b), $\frac{\sigma(e^+e^-\rightarrow \gamma*gg \rightarrow 2(s\bar{s})+(u\bar{u}+d\bar{d})/\sqrt{2})} {\sigma(e^+e^-\rightarrow \gamma*gg \rightarrow 3(s\bar{s}))}$ would range from 1 to 2.5, depending on the ratio of the two cases above.
The study of $\sigma(e^+e^-\rightarrow \phi\phi\omega)/\sigma(e^+e^-\rightarrow \phi\phi\phi)$ can therefore help to understand the production mechanism of $e^+e^-$ annihilation to multiple quarkonia.

\section{Detector and Monte Carlo Simulation}
The BESIII detector, as described in detail in Ref.~\cite{BESIII}, has a geometrical acceptance of 93\% of the solid angle. A small-cell, helium-based main drift chamber (MDC) immersed in a 1\,T magnetic field measures the momentum of charged particles with a resolution of 0.5\% at 1\,GeV/$c$, and provides energy loss (d$E$/d$x$) measurements with a resolution better than 6\% for electrons from Bhabha scattering. The electromagnetic calorimeter (EMC) detects photons with a resolution of 2.5\% (5\%) at an energy of 1\,GeV  in the barrel (end cap) region. A time-of-flight system (TOF) assists in particle identification (PID) with a time resolution of 80\,ps (110\,ps) in the barrel (end cap) region.

A {\sc geant4}-based \cite{Agostinelli:2002hh} Monte Carlo (MC) simulation software, which
includes the geometric description of the BESIII detector and
the detector response, is used to optimize the event selection
criteria, determine the detection efficiency and estimate background contributions.
The simulation includes the beam energy spread and initial-state
radiation (ISR) modeled with {\sc kkmc}~\cite{KKMC}. In this analysis, 0.5 million events of $e^+e^-\rightarrow \phi\phi\omega$ and $e^+e^-\rightarrow \phi\phi\phi$ are generated individually at different c.m.\ energies corresponding to the experimental values. Both processes are simulated with a uniform distribution in phase space (PHSP).
The observed cross sections for $e^+e^-\rightarrow \phi\phi\omega$ and $e^+e^-\rightarrow \phi\phi\phi$ at the six energy values in this analysis are used as the inputs in the KKMC simulation for ISR effects. In line with the partial reconstruction technique that is implemented in the analysis, the signal process $e^+e^-\rightarrow \phi\phi\omega$ is simulated with both $\phi$ decaying into $K^+K^-$ and the $\omega$ decaying into all possible final states, while in the simulation of $e^+e^-\rightarrow \phi\phi\phi$ events, all three $\phi$ are generated to decay via all possible modes.

\section{Event Selection}
The candidate events for $e^+e^-\rightarrow \phi\phi\omega$ and $\phi\phi\phi$ are selected with a partial reconstruction method to get higher efficiencies. We reconstruct two $\phi$ mesons with their prominent $K^+K^-$ decay mode and identify the remaining $\omega$ or $\phi$ meson with the mass recoiling against the reconstructed $\phi\phi$ system.

For each charged track, the polar angle in the MDC must satisfy $|\cos\theta|<0.93$, and the point of closest approach to the $e^{+}e^{-}$ interaction point must be within $\pm$10\,cm in the beam direction and within 1\,cm in the plane perpendicular to the beam direction. We identify charged kaon candidates using the d$E$/d$x$ and TOF information. The probabilities $\mathcal{L}(\pi)$ and $\mathcal{L}(K)$ are determined for the $\pi$ and $K$ hypothesis, respectively. Kaons are identified by requiring $\mathcal{L}(K)>\mathcal{L}(\pi)$.

The $\phi$ candidates are formed from pairs of identified kaons with opposite charges. Their invariant mass is required to satisfy $1.01<M(K^+K^-)<1.03$\,GeV/$c^2$. At least two $\phi$ candidates with no shared tracks are required in each event. If there are more than two $\phi$ candidates in one event, only the $\phi\phi$ combination with the minimum $\Delta M$ is kept for further analysis, and the two $\phi$ candidates are randomly labeled as $\phi_1$ or $\phi_2$. The mass difference $\Delta M$ is defined as $\sqrt{(M_{\phi1}(K^+K^-)-M(\phi))^{2}+(M_{\phi2}(K^+K^-)-M(\phi))^{2}}$, where $M(\phi)$ is the nominal mass of the $\phi$ meson taken from the particle data group (PDG)~\cite{PDG}.

\begin{figure*}
\begin{center}
\includegraphics[width=0.45\linewidth]{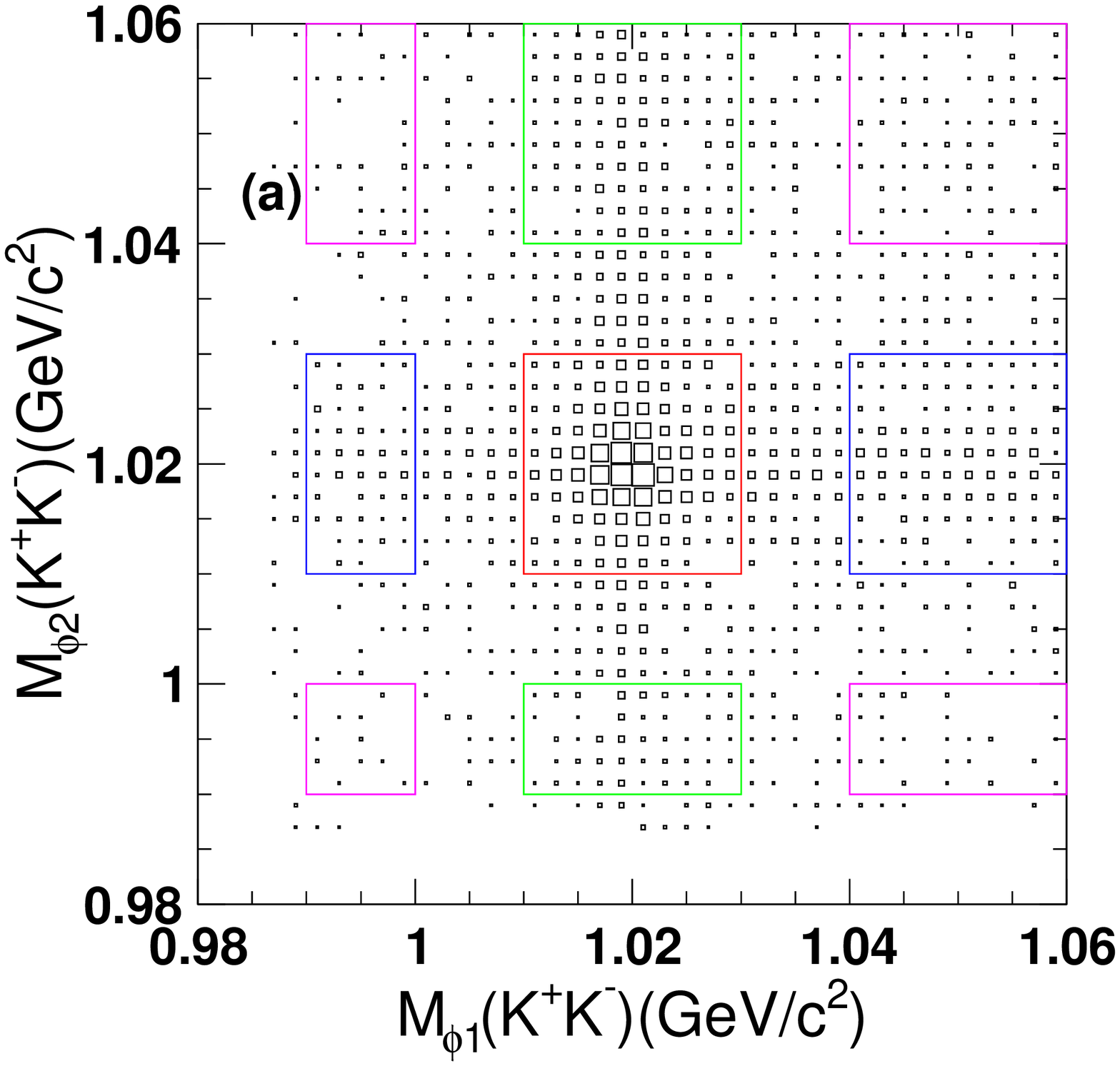}
\includegraphics[width=0.45\linewidth]{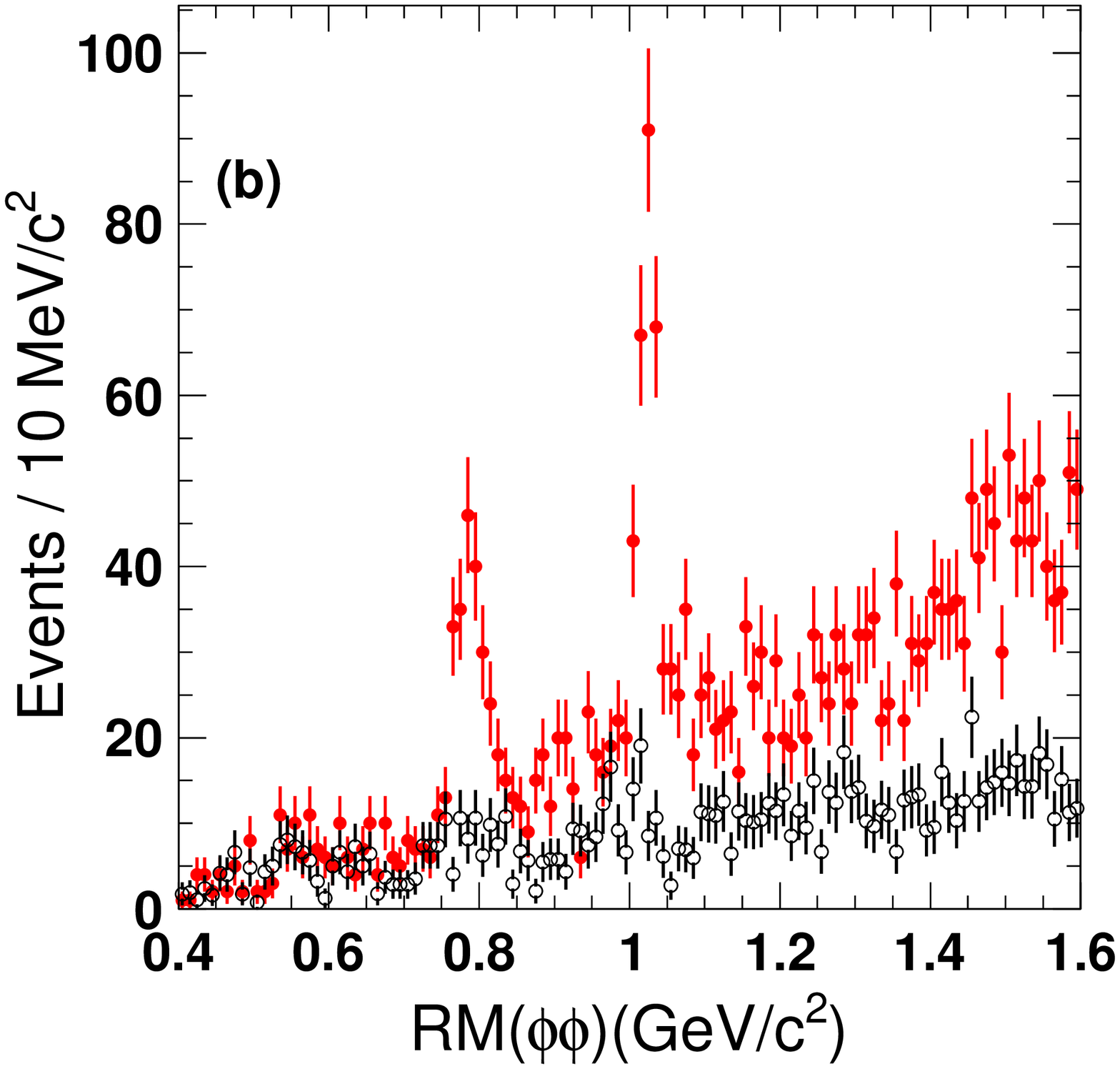}
\caption{(a) Scatter plot of $M_{\phione}(K^{+}K^{-})$ versus $M_{\phitwo}(K^{+}K^{-})$. The central box is the signal region while the boxes around are the two-dimensional sidebands. (b) The recoil mass distributions of $\phi\phi$ for events in the signal region (solid points) or sidebands (circles). All six data samples are combined.}
\label{fig:alldata-optM}
\end{center}
\end{figure*}

Fig.~\ref{fig:alldata-optM} (a) depicts the scatter plot of $M_{\phi1}(K^{+}K^{-})$ versus $M_{\phi2}(K^{+}K^{-})$ by combining the data samples at six c.m.\ energies. A clear accumulation of events is observed around the intersection of the $\phi_1$ and $\phi_2$ mass regions, which indicates $e^+e^-\rightarrow \phi\phi X$ signals. The mass of the system recoiling against the reconstructed $\phi\phi$ is calculated with $RM(\phi\phi)=\sqrt{(E_{\rm{cm}}-E_{\phi\phi})^{2}-p_{\phi\phi}^{2}}$, where $E_{\rm{cm}}$ is the c.m.\ energy obtained by analyzing the di-muon process $e^{+}e^{-}\rightarrow \gamma_\text{ISR/FSR} \mu^{+}\mu^{-}$, with a precision of 0.02\%~\cite{Ecm}. $E_{\phi\phi}$ and $p_{\phi\phi}$ are the energy and momentum of the reconstructed $\phi\phi$ pair in the $e^+e^-$ rest system. As shown by the solid points in Fig.~\ref{fig:alldata-optM} (b), we obtain two clear peaks in the vicinities of $\omega$ and $\phi$ in the $RM(\phi\phi)$ distribution, which indicates the processes $e^+e^-\rightarrow \phi\phi\omega$ and $\phi\phi\phi$, respectively.

\section{Study of Backgrounds in {\boldmath $RM(\phi\phi)$}}~\label{sec:pbkg}
To ensure that the observed $\omega$ and $\phi$ signal in the $RM(\phi\phi)$ distribution originate from the processes $e^+e^-\rightarrow \phi\phi\omega$ and $\phi\phi\phi$, we perform a study of the potential peaking backgrounds.
The two dimensional (2D) sidebands illustrated in Fig.~\ref{fig:alldata-optM}~(a) are used to study the potential background without a $\phi\phi$ pair in the final state, where the $\phi$ sidebands are defined as $0.99<M(K^+K^-)<1.00$\,GeV/$c^2$ and $1.04<M(K^+K^-)<1.06$\,GeV/$c^2$. The non-$\phi_1$ and/or non-$\phi_2$ processes are estimated by the weighted sum of the events in the horizontal and vertical sideband regions, with the entries in the diagonal sidebands subtracted to compensate for the double counting of the background without any $\phi$ in final state.
The weighting factor for the $\phi_2$ but non-$\phi_1$ events in the horizontal sidebands is the ratio of the number of $\phi_2$ but non-$\phi_1$ events under the signal region ($n_\text{bkg}^\text{sig}$) to the number of $\phi_2$ but non-$\phi_1$ events in the horizontal sidebands ($n_\text{bkg}^\text{sdb}$). $n_\text{bkg}^\text{sig}$ and $n_\text{bkg}^\text{sdb}$ are determined from the 2D fit to $M_{\phi1}(K^{+}K^{-})$ versus $M_{\phi2}(K^{+}K^{-})$. The weighting factor for the $\phi_1$ but non-$\phi_2$ (non-$\phi_1$ and non-$\phi_2$) events in the vertical (diagonal) sidebands are determined similarly.
The 2D probability density functions for the components $\phi_1\phi_2$, $\phi_1$ but non-$\phi_2$, non-$\phi_1$ but $\phi_2$, non-$\phi_1$ and non-$\phi_2$ are constructed by the product of two one-dimensional functions.
The $\phi$ peak is described with a MC-derived shape convoluted with a Gaussian function to take into account the resolution difference between data and MC simulation. The non-$\phi$ component is described with second-order polynomial functions. The estimated $RM(\phi\phi)$ distribution with weighted 2D sidebands events is shown as the open circles in Fig.~\ref{fig:alldata-optM}~(b). Since the $\phi$ signal is close to the $K^+K^-$ production threshold, we are not able to obtain a sideband which is far enough away from the signal region at the lower side of $M(K^+K^-)$. Thus, the small $\omega$ and $\phi$ signals observed in $RM(\phi\phi)$ estimated with the 2D sideband are from the leakage of the real $\ee\tto\phi\phi+\omega/\phi$ signals.
From studies of signal MC samples, the ratio of the signal events in the 2D sideband regions to those in the signal region is estimated to be 3\%$\sim$5\%.

We also estimate the peaking background in the $RM(\phi\phi)$ distribution for the process $e^+e^-\rightarrow \phi \phi \phi$ with the MC samples. The dominant peaking backgrounds is from the \mbox{$e^+e^-\rightarrow K^+K^- \phi\phi$} and \mbox{$e^+e^-\rightarrow K^+K^-K^+K^-\phi$} processes. When the directly produced $K^+K^-$ ($K^+K^-K^+K^-$) is reconstructed as $\phi$ ($\phi\phi$), these two processes would contribute as peaking backgrounds in the $RM(\phi\phi)$ distribution. 
The contamination rate of the \mbox{$e^+e^-\rightarrow K^+K^- \phi\phi$} (\mbox{$e^+e^-\rightarrow K^+K^-K^+K^- \phi$}) events to $e^+e^-\rightarrow \phi \phi \phi$ is estimated to be $\sim1.0$\% (0.1\%) at each energy point with the assumption that the c.m.\ energy dependent cross section for \mbox{$e^+e^-\rightarrow K^+K^- \phi\phi$} (\mbox{$e^+e^-\rightarrow K^+K^-K^+K^- \phi$}) is the same as for $e^+e^-\rightarrow \phi \phi \phi$.
We take 1.0\% as the uncertainty on the size of the peaking backgrounds of $e^+e^-\rightarrow \phi \phi \phi$.
Similarly, the dominant peaking backgrounds of $e^+e^-\rightarrow \phi \phi \omega$ is from the \mbox{$e^+e^-\rightarrow K^+K^- \phi\omega$} and \mbox{$e^+e^-\rightarrow K^+K^-K^+K^-\omega$} processes. For $e^+e^-\rightarrow \phi \phi \omega$, the uncertainty from the peaking backgrounds is determined to be 1.0\%.

\section{Fits to the $RM(\phi\phi)$ Spectrum and Cross Section Results}
The reconstruction efficiencies and yields of $\ee\tto\phi\phi\omega$ and $\phi\phi\phi$ signals are determined by the fit to the $RM(\phi\phi)$ distribution for MC simulation and data, respectively.

\subsection{Correction to $RM(\phi\phi)$}
Compared with the values in the PDG, the measured masses of the $\omega$ and $\phi$ mesons in the $RM(\phi\phi)$ distribution deviate to the left with $\sim$4.5\,MeV. 
This deviation may be induced by ISR, the energy loss of the reconstructed kaons and final state radiation (FSR), or the uncertainty of $E_{\rm{cm}}$. The overall effect is considered as a shift on  $E_{\rm{cm}}$, $\Delta E_{\rm{cm}}$.

We estimate $\Delta E_{\rm{cm}}$ by studying the process $e^+e^-\rightarrow \phi K^+ K^-$ with partially reconstructing one $\phi$ meson and one charged kaon. The recoil mass against the reconstructed $\phi K$ is calculated with $RM(\phi K)=\sqrt{(E_{\rm{cm}}-E_{\phi K})^{2}-p_{\phi K}^{2}}$, where $E_{\phi K}$ and $p_{\phi K}$ are the energy and momentum of the reconstructed $\phi K$ in the system of $e^+e^-$. $\Delta E_{\rm{cm}}$ is estimated with \mbox{$\Delta E_{\rm{cm}}=\frac{RM(\phi K)}{E_{\rm{cm}}-E_{\phi K}} \times \Delta RM(\phi K)$}, where $RM(\phi K)$ is approximately $m(K)$ from PDG and $E_{\phi K}$ is the average over all $\phi K^+ K^-$ events. $RM(\phi\phi)$ for each event is then corrected by subtracting $\Delta RM(\phi\phi)$ in the data and MC samples, where \mbox{$\Delta RM(\phi\phi)=\frac{E_{\rm{cm}}-E_{\phi\phi}}{RM(\phi\phi)} \times \Delta E_{\rm{cm}}$}. As a consequence, the measured masses of the $\omega$ and $\phi$ mesons obtained by fitting the $RM(\phi\phi)$ distributions are consistent with the values in the PDG.

\begin{figure*}
\begin{center}
\includegraphics[width=0.9\linewidth]{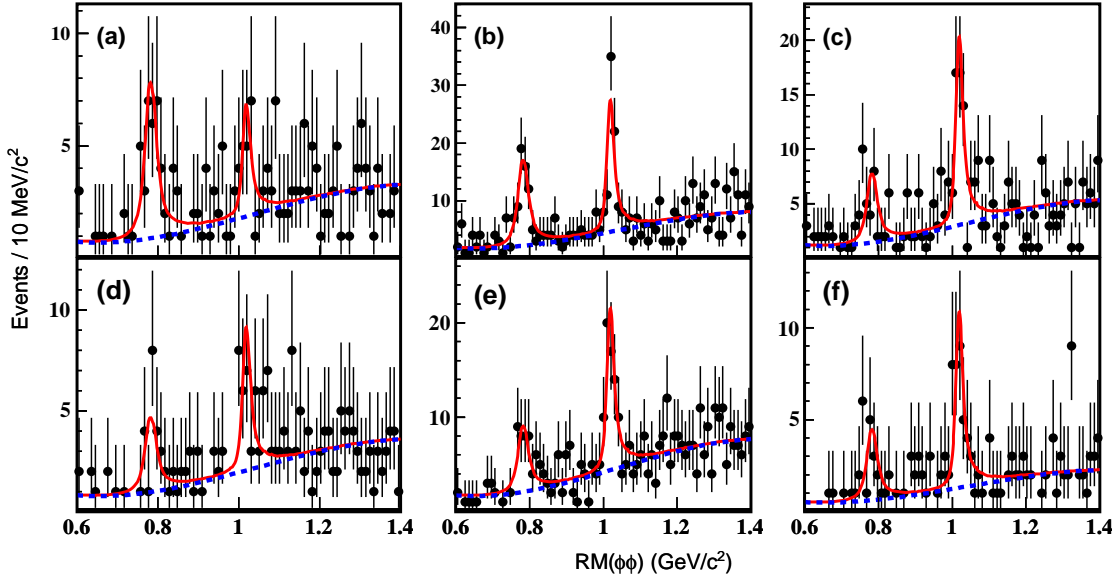}
\caption{Fits to the corrected $RM(\phi\phi)$ distribution for data samples at $E_{\rm{cm}}=$ (a) 4.008, (b) 4.226, (c) 4.258, (d) 4.358, (e) 4.416 and (f) 4.600 \,GeV. In each plot, the points with error bar are data, the dashed curve is the background contribution and the solid line shows the total fit.}
\label{fig:Yields}
\end{center}
\end{figure*}

\subsection{Fits to the $RM(\phi\phi)$ Spectrum}
An unbinned maximum likelihood fit is performed to the corrected $RM(\phi\phi)$ distributions. The signal distribution is modeled by the MC-derived signal shape.
The study of the selected $\phi$ signal indicates that the mass resolution difference for the $\phi$ signal is very small. Therefore, we assume the resolution of $RM(\phi\phi)$ is the same between data and MC simulation, and the corresponding systematic uncertainty will be considered. The background shape is described by a third-order Chebyshev polynomial function with parameters fixed to the values obtained by fitting all samples together, since some samples have small statistics. The corresponding fit results are shown in Fig.~\ref{fig:Yields}. The statistical significances of the $\omega/\phi$ signals are examined using the differences in likelihood values of fits with and without an $\omega/\phi$ signal component included in the fits. Both $\omega$ and $\phi$ signals are seen with statistical significances of more than 3$\sigma$ for each data sample, and the significances of $\omega$ and $\phi$ are both larger than 10\,$\sigma$ if all six data samples are combined. The yields of $\omega$ and $\phi$ signal events and the corresponding statistical significances for each sample are summarized in Table~\ref{table:cs-ppox-optM} and Table~\ref{table:cs-pppx-optM}, respectively.

\subsection{Reconstruction Efficiency}
\begin{figure*}
\begin{center}
\includegraphics[width=0.9\linewidth]{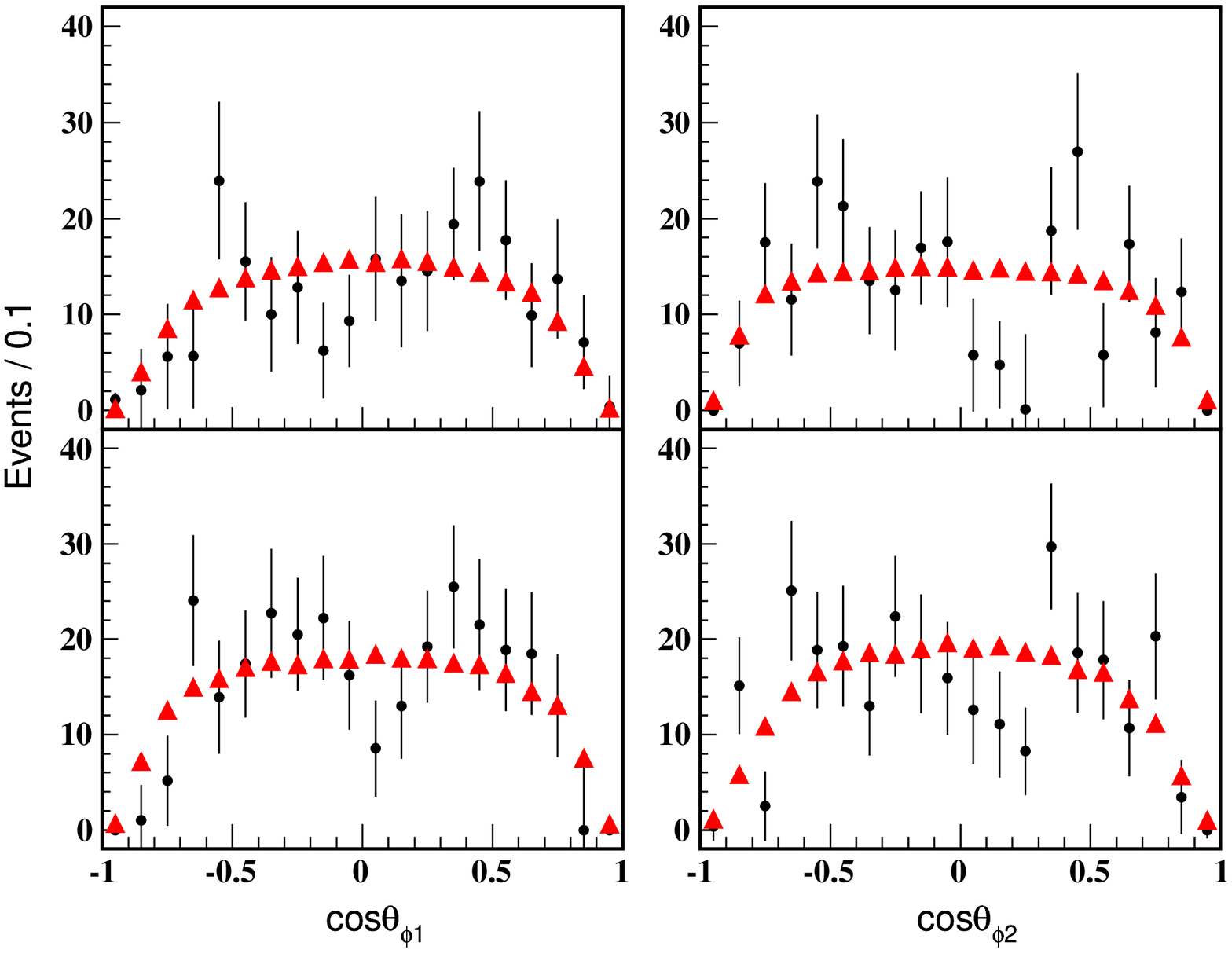}
\caption{Comparison of the $\cos{\theta}$ distributions in data (points) and PHSP MC simulation (triangles), for $\ee\tto\phi\phi\omega$ (top plots) and $\ee\tto\phi\phi\phi$ (bottom plots) signals, combining all data samples. The $\cos{\theta}$ distributions are obtained by fitting the $RM(\phi\phi)$ distribution for events with  $\cos{\theta}$ in given bins.}
\label{fig:comp-cos-kpall-PPPX}
\end{center}
\end{figure*}

The $\ee\tto\phi\phi\omega$ and $\phi\phi\phi$ signal MC samples are simulated by assuming a uniform distribution in phase space. The reconstruction efficiency of the two reconstructed $\phi$s depends on their production angles. The comparison of the cosine of the polar angles  $\theta$ for the two reconstructed $\phi$ mesons between data and MC simulation is presented in Fig.~\ref{fig:comp-cos-kpall-PPPX}, where the $\cos{\theta}$ distributions are obtained by fitting the $RM(\phi\phi)$ distribution for events with  $\cos{\theta}$ in given bins. All the data samples are combined, assuming the $\cos{\theta}$ distributions do not depend on the c.m.\ energy.
To take into account the deviation in $\cos{\theta}$ distributions between the data and the PHSP MC samples, the reconstruction efficiencies are determined with PHSP MC samples incorporating the re-weighting correction according to the 2D distribution of $\cos{\theta_1}$ versus $\cos{\theta_2}$ of data and PHSP MC samples.

\begin{table*}[]\footnotesize
\renewcommand{\tablename}{TABLE}
\begin{center}
\caption{Summary of the measurements of the $\ee\tto\phi\phi\omega$ process. Listed in the table are the c.m. energy $E_{\rm{cm}}$, the integrated luminosity $\mathcal{L}_{\rm{int}}$, the number of the observed events $N^{\rm{obs}}$, the reconstruction efficiency $\epsilon$, the vacuum polarization factor $(1+\delta^v)$, the radiative correction factor $(1+\delta^r)$, the measured Born cross section $\sigma^{\rm{B}}$, and statistical significance. The first uncertainty of the Born cross section is statistical, and the second is systematic.}
\label{table:cs-ppox-optM}
\begin{tabular}{c|c|c|c|c|c|c|c}
\hline\hline
$E_{\rm{cm}}$(GeV)&$\mathcal{L}_{\rm{int}}(\rm{pb}^{-1})$ & $N^{\rm{obs}}$ & $\epsilon$($\%$) &  $(1+\delta^v)$ & $(1+\delta^r)$ & $\sigma^{\rm{B}}$(fb) & Significance\\  \hline
4.008&482.0&36.0$\pm$7.6&22.7&1.044&0.888&1485$\pm$312$\pm$138&7.3$\sigma$\\
4.226&1091.7&82.6$\pm$11.8&25.3&1.057&0.940&1260$\pm$180$\pm$94&10.6$\sigma$\\
4.258&825.7&41.0$\pm$9.6&25.2&1.054&1.159&674$\pm$158$\pm$56&5.8$\sigma$\\
4.358&539.8&23.5$\pm$7.1&25.8&1.051&1.062&633$\pm$191$\pm$47&4.6$\sigma$\\
4.416&1073.6&44.1$\pm$10.1&25.6&1.053&1.054&605$\pm$138$\pm$50&5.9$\sigma$\\
4.600&566.9&24.1$\pm$6.6&26.3&1.055&0.995&643$\pm$177$\pm$50&5.3$\sigma$\\
\hline
\end{tabular}
\end{center}
\end{table*}

\begin{table*}[]\footnotesize
\renewcommand{\tablename}{TABLE}
\begin{center}
\caption{Summary of the measurements of the $\ee\tto\phi\phi\phi$ process. Listed in the table are the c.m. energy $E_{\rm{cm}}$, the number of the observed events $N^{\rm{obs}}$, the reconstruction efficiency $\epsilon$, the radiative correction factor $(1+\delta^r)$, the measured Born cross section $\sigma^{\rm{B}}$, and statistical significance. The first uncertainty of the Born cross section is statistical, and the second is systematic. The integrated luminosity $\mathcal{L}_{\rm{int}}$ and the vacuum polarization factor $(1+\delta^v)$ are same with those in Table~\ref{table:cs-ppox-optM}.}
\label{table:cs-pppx-optM}
\begin{tabular}{c|c|c|c|c|c}
\hline\hline
$E_{\rm{cm}}$(GeV) & $N^{\rm{obs}}$ & $\epsilon$($\%$) & $(1+\delta^r)$ & $\sigma^{\rm{B}}$(fb) & Significance\\  \hline
4.008&17.9$\pm$6.5&59.8&0.876&284$\pm$104$\pm$28&3.5$\sigma$\\
4.226&82.6$\pm$12.1&68.3&0.876&500$\pm$73$\pm$55&9.7$\sigma$\\
4.258&63.9$\pm$10.8&69.2&0.886&501$\pm$85$\pm$56&8.4$\sigma$\\
4.358&31.2$\pm$8.8&70.4&0.983&332$\pm$94$\pm$40&4.6$\sigma$\\
4.416&68.4$\pm$11.9&71.6&0.932&379$\pm$66$\pm$45&7.7$\sigma$\\
4.600&39.2$\pm$8.2&73.7&0.942&395$\pm$83$\pm$49&6.9$\sigma$\\
\hline
\end{tabular}
\end{center}
\end{table*}

\begin{figure*}
\begin{center}
\includegraphics[width=0.32\linewidth]{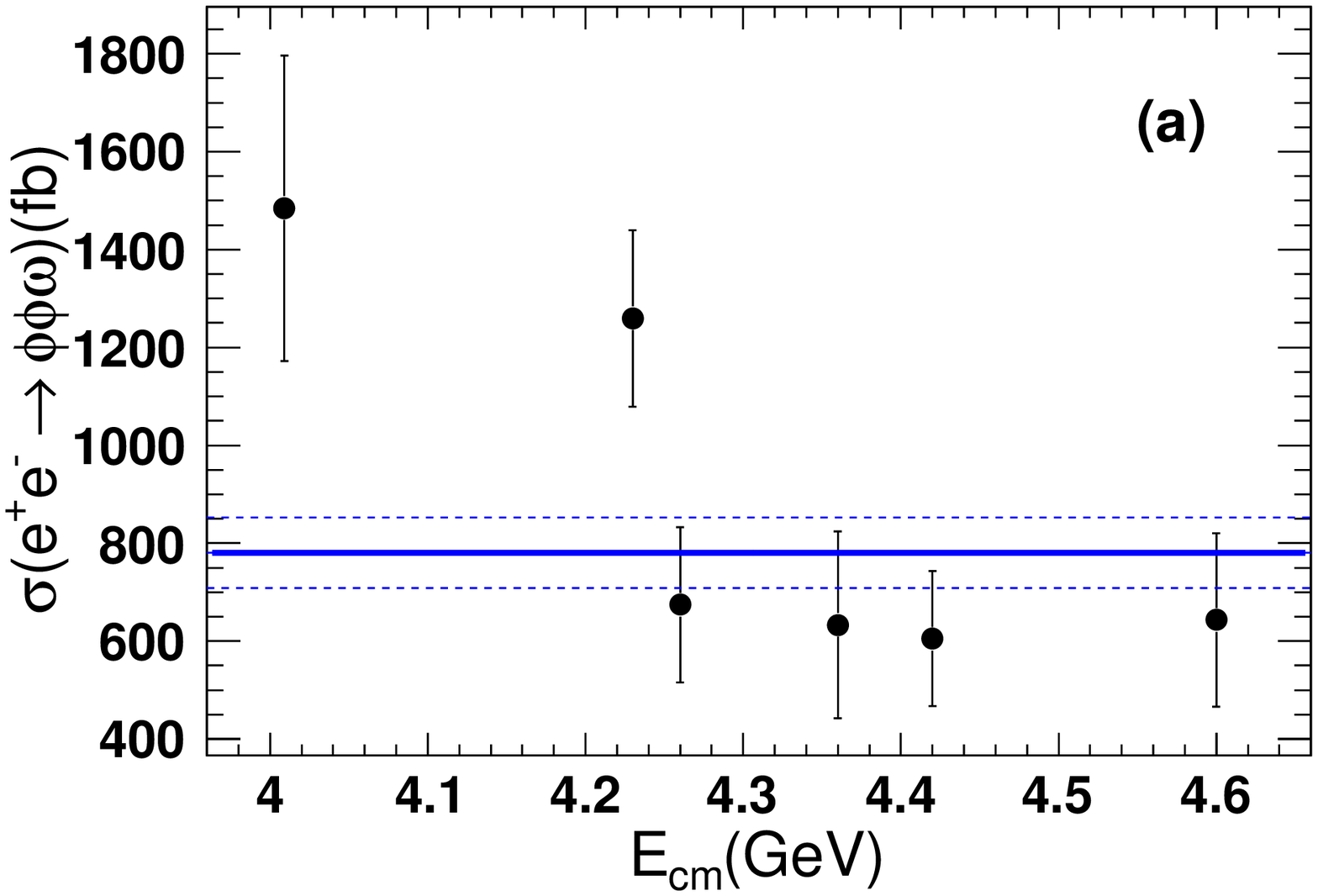}
\includegraphics[width=0.32\linewidth]{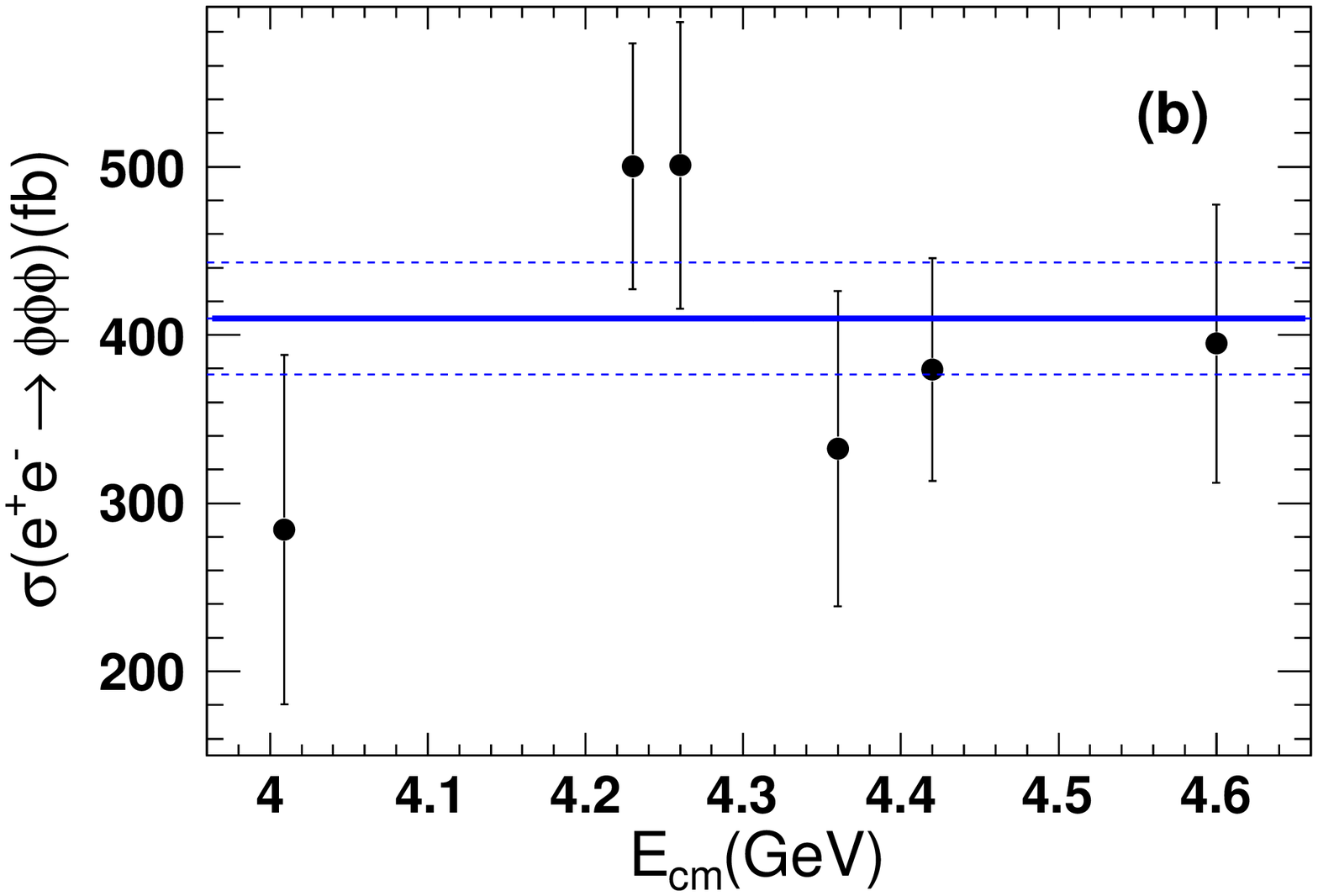}
\includegraphics[width=0.32\linewidth]{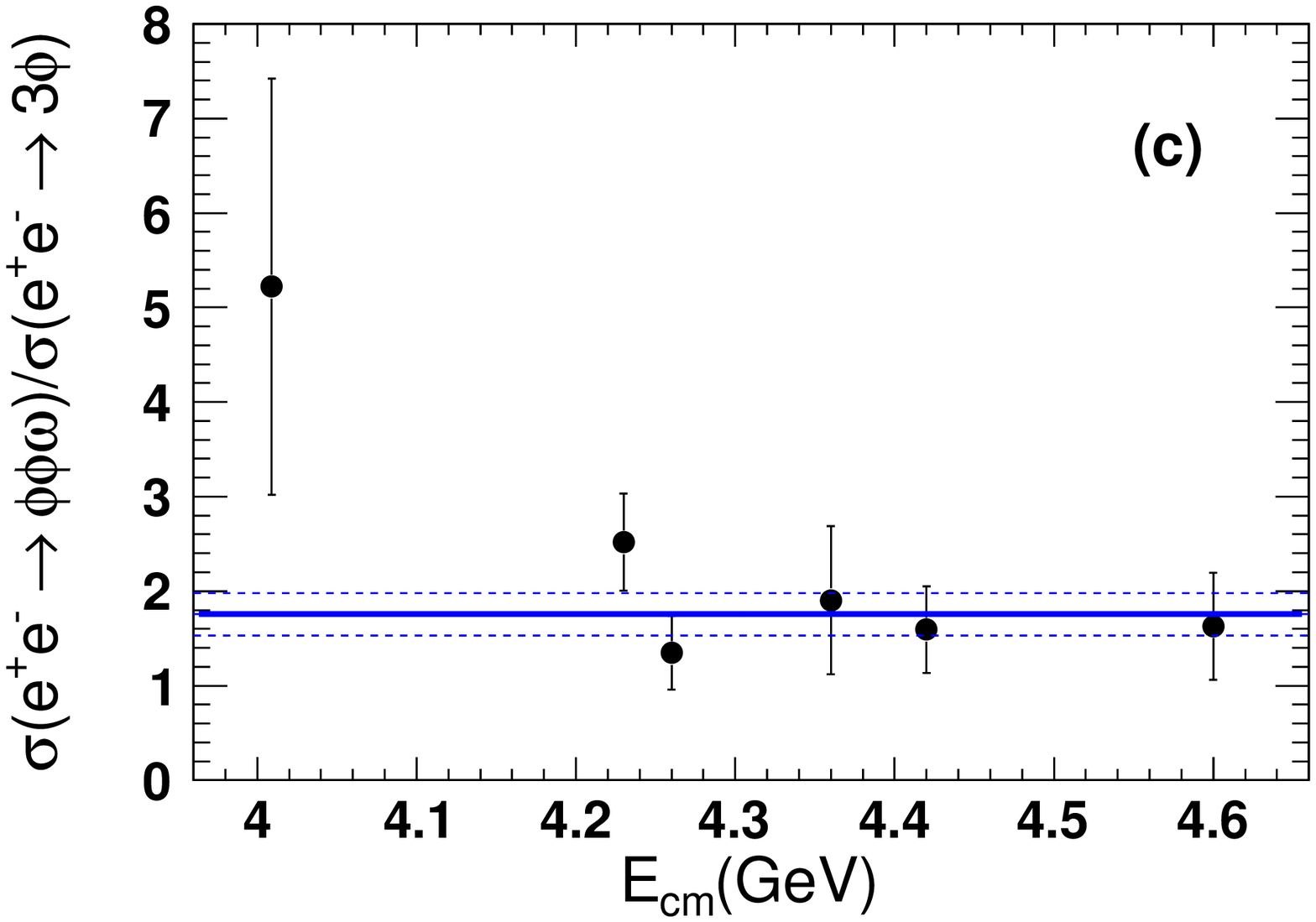}
\caption{Born cross sections of (a) $\ee\tto\phi\phi\omega$ and (b) $\ee\tto\phi\phi\phi$ at six energy points. (c) Ratios $\sigma(\ee\tto\phi\phi\omega)/\sigma(\ee\tto \phi\phi\phi)$. The blue lines show the statistical-weighted averages with an error band corresponding to one standard deviation of the statistical uncertainty.}
\label{fig:cs-optM}
\end{center}
\end{figure*}
\subsection{Cross Section Results}
The Born cross section is calculated by
\begin{equation}
\sigma^{\rm{B}}=\frac{N^{\rm{obs}}}{\mathcal{L}_{\rm{int}}\cdot(1+\delta^{r})\cdot(1+\delta^{v})\cdot\epsilon\cdot\mathcal{B}^2}
\label{eq:cs}
\end{equation}
where $N^{\rm{obs}}$ is the number of observed signal events, $\mathcal{L}_{\rm{int}}$ is the integrated luminosity, $(1+\delta^{r})$ is the radiative correction factor, $(1+\delta^{v})$ is the vacuum polarization factor, $\epsilon$ is the detection efficiency including reconstruction and all selection criteria, and $\mathcal{B}$ is the branching fraction of $\phi\rightarrow K^+K^-$. The vacuum polarization factor is taken from a QED calculation.
With the input of the observed c.m.\ energy dependent $\sigma(\ee\tto\phi\phi\omega)$ and $\sigma(\ee\tto\phi\phi\phi)$, and using a linear interpolation to obtain the cross sections in the full range, the radiative correction factor is calculated in QED~\cite{delta-r}. Since the radiative correction factor and the detection efficiency both depend on the line shape of the input cross section, the Born cross sections of $\ee\tto\phi\phi\omega$ and $\ee\tto\phi\phi\phi$ are determined with four iterations until convergence has been reached. The values of all variables used in the calculation of $\sigma(e^+e^-\rightarrow \phi\phi\omega)$ and $\sigma(e^+e^-\rightarrow\phi\phi\phi)$ are listed in Table~\ref{table:cs-ppox-optM} and Table~\ref{table:cs-pppx-optM}, respectively.

Fig.~\ref{fig:cs-optM} (a) and (b) show the measured Born cross sections $\sigma(\ee\tto\phi\phi\omega)$ and $\sigma(\ee\tto\phi\phi\phi)$, respectively. The statistical-weighted average of the measurements at different c.m.\ energies is shown as the flat line. Variations within one standard deviation of the statistical uncertainty are shown with the dashed lines. The measured Born cross sections of $\ee\tto\phi\phi\phi$ are compatible with a flat distribution, with $\chi^2/DOF=5.1/5$, while for the $\ee\tto\phi\phi\omega$ process the compatibility is poor with $\chi^2/DOF=15.4/5$.

\begin{table*}[]\small
\caption{Summary of systematic uncertainties (\%) in the measurement of $\sigma(e^+e^-\rightarrow \phi\phi\omega)$.}
\centering
\begin{tabular}{c|ccccccccc|c}
\hline
\hline
\multirow{2}{*}{$E_{\rm{cm}}$(GeV)} & \multirow{2}{*}{Tracking} & \multirow{2}{*}{PID} & Background   & Peaking & Line & \multirow{2}{*}{$\delta^{v}$} & Simulation & \multirow{2}{*}{$\mathcal{L}_{\rm{int}}$} &
\multirow{2}{*}{$\mathcal{B}$}  & \multirow{2}{*}{Total} \\
     &     &   &  shape  & backgrounds  &  shape & &  model  &  &  &            \\ \hline
 4.008  & 4.0 & 4.0 & 1.9 & 1.0 & 0.9 & 0.5 & 6.6 & 1.0 & 2.0 & 9.3 \\
 4.226  & 4.0 & 4.0 & 2.3 & 1.0 & 0.5 & 0.5 & 3.8 & 1.0 & 2.0 & 7.5 \\
 4.258  & 4.0 & 4.0 & 3.7 & 1.0 & 0.6 & 0.5 & 4.2 & 1.0 & 2.0 & 8.3 \\
 4.358  & 4.0 & 4.0 & 2.5 & 1.0 & 0.5 & 0.5 & 3.4 & 1.0 & 2.0 & 7.4 \\
 4.416  & 4.0 & 4.0 & 3.4 & 1.0 & 0.2 & 0.5 & 4.1 & 1.0 & 2.0 & 8.2 \\
 4.600  & 4.0 & 4.0 & 2.5 & 1.0 & 3.4 & 0.5 & 2.6 & 1.0 & 2.0 & 7.8 \\
\hline
\hline
\end{tabular}
\label{table:syserr-omg}
\end{table*}

\begin{table*}[]\small
\caption{Summary of systematic uncertainties (\%) in the measurement of $\sigma(e^+e^-\rightarrow \phi\phi\phi)$.}
\centering
\begin{tabular}{c|ccccccccc|c}
\hline
\hline
\multirow{2}{*}{$E_{\rm{cm}}$(GeV)} & \multirow{2}{*}{Tracking} & \multirow{2}{*}{PID} & Background   & Peaking & Line & \multirow{2}{*}{$\delta^{v}$} & Simulation & \multirow{2}{*}{$\mathcal{L}_{\rm{int}}$} &
\multirow{2}{*}{$\mathcal{B}$}  & \multirow{2}{*}{Total} \\
     &     &   &  shape  & backgrounds  &  shape  & & model  & &   &            \\ \hline
 4.008  & 4.0 & 4.0 & 3.7 & 1.0 & 0.1 & 0.5  & 7.0 & 1.0 & 2.0 & 10.0 \\
 4.226  & 4.0 & 4.0 & 1.9 & 1.0 & 0.8 & 0.5  & 8.8 & 1.0 & 2.0 & 10.9 \\
 4.258  & 4.0 & 4.0 & 2.0 & 1.0 & 1.5 & 0.5  & 9.1 & 1.0 & 2.0 & 11.2 \\
 4.358  & 4.0 & 4.0 & 2.9 & 1.0 & 0.8 & 0.5  & 9.8 & 1.0 & 2.0 & 11.9 \\
 4.416  & 4.0 & 4.0 & 2.4 & 1.0 & 2.6 & 0.5  & 9.6 & 1.0 & 2.0 & 11.9 \\
 4.600  & 4.0 & 4.0 & 1.5 & 1.0 & 2.7 & 0.5  & 10.2 & 1.0 & 2.0 & 12.3 \\
\hline
\hline
\end{tabular}
\label{table:syserr-phi}
\end{table*}

\section{Systematic Uncertainties of Cross Sections}
Several sources of systematic uncertainties are considered in the measurement of the Born cross sections. These include differences between the data and the MC simulation for the tracking efficiency, PID efficiency, mass window requirement, the MC simulation of the radiative correction factor and the vacuum polarization factor. We also consider the uncertainties from the fit procedure, the peaking backgrounds, the simulation model as well as uncertainties of the branching fraction of $\phi\rightarrow K^+K^-$ and the integrated luminosity.

\begin{enumerate}[a.]
\item {\it Tracking efficiency.}~~~The difference in tracking efficiency for the kaon reconstruction between the data and the MC simulation is estimated to be 1.0\% per track~\cite{Tracking}. Therefore, 4.0\% is taken as the systematic uncertainty for four kaons.
\item {\it PID efficiency.}~~~PID is required for the kaons, and the uncertainty is estimated to be 1.0\% per kaon~\cite{Tracking}. Hence, 4.0\% is taken as the systematic uncertainty of the PID efficiency for four kaons.
\item {\it $\phi$ mass window.}~~~A mass window requirement on the $K^+K^-$ invariant mass might introduce a systematic uncertainty on the efficiency. The reconstructed $\phi$ signals are fit with a MC shape convoluted with a Gaussian function that describes the disagreement between data and MC simulation. The mean and width of the Gaussian function are left free in the fit, which turn out to be close to 0 within 3 times of uncertainty. The systematic uncertainty from the $M(K^+K^-)$ requirement is ignored.
\item {\it Fit procedure.}~~~For the six data samples, the yields of $e^+e^-\rightarrow \phi\phi\omega$ and $\phi\phi\phi$ events are obtained by a fit to the distribution of the mass recoiling against the reconstructed $\phi\phi$ system. The following two aspects are considered when evaluating the systematic uncertainty associated with the fit procedure. (1) \emph{Signal shape.}---In the nominal fit, the signal shapes are described by the MC shape obtained from MC simulation. An alternative fit with the MC shape convoluted with a Gaussian function for the $\omega/\phi$ signal shape is performed, where the parameters of the Gaussian function are free. The resulting difference in the yield with respect to the nominal fit is considered as the systematic uncertainty from the signal shape. This uncertainty is negligible compared to the statistical uncertainty. (2) \emph{Background shape.}---In the nominal fit, the background shape is described with a third-order Chebyshev polynomial function. The fit with a fourth-order Chebyshev polynomial function for the background shape is performed to estimate the uncertainty due to the background parametrization.
\item {\it Peaking backgrounds.}~~~The uncertainty is taken as 1.0\%, as described in Sec.~\ref{sec:pbkg}.
\item {\it Line shape of cross section.}~~~The line shape of the $e^+e^-\rightarrow \phi\phi\omega$ and $\phi\phi\phi$ cross sections affects the radiative correction factor and the reconstruction efficiency. The corresponding uncertainty is estimated by changing the input of the observed line shape within one standard deviation.
\item {\it vacuum polarization factor.} The QED calculation used to determine the vacuum polarization factor has an accuracy of 0.5\%~\cite{delta-v}.
\item {\it Simulation model.}~~~The differences between the efficiencies obtained with and without re-weighting the PHSP MC sample are taken as the uncertainties associated with the simulation model.
\item {\it Luminosity.}~~~The time-integrated luminosity~\cite{Gaoq-lum} of each sample is measured with a precision of 1\% with Bhabha events.
\item {\it Branching fractions.}~~~The uncertainty in the branching fraction for the process $\phi\rightarrow K^+K^-$ is taken from the PDG~\cite{PDG}.
\end{enumerate}

Assuming all of the systematic uncertainties shown in Tables~\ref{table:syserr-omg} and~\ref{table:syserr-phi} are independent, the total systematic uncertainties are obtained by adding the individual uncertainties in quadrature.

\begin{table*}[]\footnotesize
\renewcommand{\tablename}{TABLE}
\begin{center}
\caption{Summary of the measured $r_{\rm{cs}}$ at different c.m.\ energies and the statistical-weighted average over all samples. The first uncertainty is statistical, and the second is systematic.}
\label{table:cs-ratio-optM}
\begin{tabular}{c|c|c}
\hline\hline
$E_{\rm{cm}}$(GeV)&$r_{\rm{cs}}$&Averaged $r_{\rm{cs}}$\\  \hline
4.008&5.22$\pm$2.20$\pm$0.55& \multirow{6}{*}{1.75$\pm$0.22$\pm$0.19} \\
4.226&2.52$\pm$0.51$\pm$0.25& \\
4.258&1.35$\pm$0.39$\pm$0.15& \\
4.358&1.90$\pm$0.79$\pm$0.21& \\
4.416&1.59$\pm$0.46$\pm$0.18& \\
4.600&1.63$\pm$0.56$\pm$0.19& \\
\hline
\end{tabular}
\end{center}
\end{table*}
\section{Ratio $\sigma(e^+e^-\rightarrow \phi\phi\omega)/\sigma(e^+e^-\rightarrow \phi\phi\phi)$}
The right plot of Fig.~\ref{fig:cs-optM} shows the measured ratios  $r_{\rm{cs}}\equiv\sigma(\ee\tto\phi\phi\omega)$/$\sigma(\ee\tto\phi\phi\phi)$ at different c.m.\ energy, and the statistical-weighted average. Except for the measurement at 4.008\,GeV, the ratios are consistent with each other within one statistical standard deviation. In the calculation of $r_{\rm{cs}}$, many uncertainties on the cross sections cancel, such as the uncertainties in the tracking, PID, $\mathcal{B}(\phi\rightarrow K^+K^-)$ and luminosity. Only the uncertainties from the background shape, line shape and MC simulation model are considered in the determination of $r_{\rm{cs}}$. From the measurements at six energy points in Table~\ref{table:cs-ratio-optM}, we obtain the statistical-weighted average $r_{\rm{cs}}=1.75\pm0.22\pm0.19$, where the first uncertainty is statistical and the second systematic. The systematic uncertainties of $r_{\rm{cs}}$ at different c.m.\ energies are assumed to be independent in this calculation.

\section{Summary and Discussion}
With the data samples collected between 4.008 and 4.600\,GeV with the BESIII detector, the processes $\ee\tto\phi\phi\omega$ and $\ee\tto \phi\phi\phi$ are observed for the first time. The Born cross sections are determined at six c.m.\ energies and the average ratio $\sigma(\ee\tto\phi\phi\omega)/\sigma(\ee\tto \phi\phi\phi)$ over the six c.m.\ energies is calculated to be $1.75\pm0.22\pm0.19$, which is in the range of the estimation with Fig.~\ref{fig:FMD-phiphiX}.
Our measurements of these two processes provide experimental constraints on the theoretical calculations of the three vectors production in the $e^+e^-$ annihilation.

\section{Acknowledgement}
The BESIII collaboration thanks the staff of BEPCII and the IHEP computing center for their strong support. This work is supported in part by National Key Basic Research Program of China under Contract No. 2015CB856700; National Natural Science Foundation of China (NSFC) under Contracts Nos. 11235011, 11335008, 11425524, 11625523, 11635010, 11175189; the Chinese Academy of Sciences (CAS) Large-Scale Scientific Facility Program; the CAS Center for Excellence in Particle Physics (CCEPP); Joint Large-Scale Scientific Facility Funds of the NSFC and CAS under Contracts Nos. U1332201, U1532257, U1532258; CAS under Contracts Nos. KJCX2-YW-N29, KJCX2-YW-N45, QYZDJ-SSW-SLH003; 100 Talents Program of CAS; National 1000 Talents Program of China; INPAC and Shanghai Key Laboratory for Particle Physics and Cosmology; German Research Foundation DFG under Contracts Nos. Collaborative Research Center CRC 1044, FOR 2359; Istituto Nazionale di Fisica Nucleare, Italy; Joint Large-Scale Scientific Facility Funds of the NSFC and CAS; Koninklijke Nederlandse Akademie van Wetenschappen (KNAW) under Contract No. 530-4CDP03; Ministry of Development of Turkey under Contract No. DPT2006K-120470; National Natural Science Foundation of China (NSFC) under Contract No. 11505010; National Science and Technology fund; The Swedish Resarch Council; U. S. Department of Energy under Contracts Nos. DE-FG02-05ER41374, DE-SC-0010118, DE-SC-0010504, DE-SC-0012069; University of Groningen (RuG) and the Helmholtzzentrum fuer Schwerionenforschung GmbH (GSI), Darmstadt; WCU Program of National Research Foundation of Korea under Contract No. R32-2008-000-10155-0.

\end{multicols}

\section{References}
\begin{thebibliography}{**}
\bibitem{Bes3-R}
J.~Z.~Bai {\it et al.}, BES Collaboration, Phys.\ Rev.\ Lett. {\bf 84} (2000) 594.\\
J.~Z.~Bai {\it et al.}, BES Collaboration, Phys.\ Rev.\ Lett. {\bf 88} (2002) 101802.\\
M.~Ablikim {\it et al.}, BES Collaboration, Phys.\ Lett.\ B {\bf 677} (2009) 239.

\bibitem{exh-Babar}
B.~Aubert {\it et al.}, BaBar Collaboration, Phys.\ Rev.\ D {\bf 77} (2008) 119902.\\
B.~Aubert {\it et al.}, BaBar Collaboration, Phys.\ Rev.\ D {\bf 77} (2008) 092002.

\bibitem{exh-BESIII}
M.~Ablikim {\it et al.}, BESIII Collaboration, Phys.\ Lett.\ B {\bf 753} (2016) 629.\\
M.~Ablikim {\it et al.}, BESIII Collaboration, Phys.\ Rev.\ D {\bf 91} (2015) 11.

\bibitem{KKMC}
S.~Jadach, B.F.L.~Ward, Z.~Was, Phys.\ Rev.\  D {\bf 63} (2001) 113009.

\bibitem{Pythia}
T.~Sjostrand, L.~Lonnblad, S.~Mrenna, P.~Skands, hep-ph/0108264, LU TP 01-21 (2002).\\
http://home.thep.lu.se/~torbjorn/pythiaaux/past.html

\bibitem{Belle-jpsiccbar}
K.~Abe {\it et al.}, Belle Collaboration, Phys.\ Rev.\ D {\bf 70} (2004) 071102. \\
B.~Aubert {\it et al.}, BABAR Collaboration, Phys.\ Rev.\ D {\bf 72} (2005) 031101.

\bibitem{NRQCD}
G.~T.~Bodwin, E.~Braaten, and G.~P.~Lepage, Phys.\ Rev.\ D {\bf 51} (1995) 1125.

\bibitem{NLO}
Y.-J.~Zhang, Y.-j.~Gao, and K.-T.~Chao, Phys.\ Rev.\ Lett. {\bf 96} (2006) 092001. \\
B.~Gong and J.-X.~Wang, Phys.\ Rev.\ D {\bf 77} (2008) 054028.

\bibitem{pred-Jpsiccbar}
E.~Braaten and J.~Lee, Phys.\ Rev.\ D {\bf 67} (2003) 054007; Phys.\ Rev.\ D {\bf 72} (2005) 099901(E). \\
K.-Y. Liu, Z.-G. He, and K.-T. Chao, Phys.\ Lett.\ B {\bf 557} (2003) 45. \\
K.~Hagiwara, E.~Kou, and C.-F.~Qiao, Phys.\ Lett.\ B {\bf 570} (2003) 39.

\bibitem{Ecm}
M.~Ablikim {\it et al.}, BESIII Collaboration, Chin.\ Phys.\ C {\bf 40} (2016) 063001.

\bibitem{BEPCII}
F. A. Harris, Nucl.\ Phys.\ Proc.\ Suppl. {\bf 162} (2006) 345.

\bibitem{BESIII}
M.~Ablikim {\it et al.}, BESIII Collaboration, Nucl.\ Instrum.\ Meth.\ A {\bf 614} (2010) 345.

\bibitem{Agostinelli:2002hh}
S.~Agostinelli {\it et al.}, GEANT4 Collaboration, Nucl.\ Instrum.\ Meth.\ A {\bf 506} (2003) 250.

\bibitem{PDG}
C.~Patrignani {\it et al.}, Particle Data Group, Chin.\ Phys.\ C {\bf 40} (2016) 100001.

\bibitem{delta-r}
E.~A.~Kuraev and V.~S.~Fadin, Sov.\ J.\ Nucl.\ Phys. {\bf 41} (1985) 466 [Yad.\ Fiz {\bf 41} (1985) 733].

\bibitem{Tracking}
M.~Ablikim {\it et al.}, BESIII Collaboration, Phys.\ Rev.\ Lett. {\bf 112} (2014) 022001.

\bibitem{delta-v}
S.~Actis {\it et al.} Eur.\ Phys.\ J.\ C {\bf 66} (2010) 585.

\bibitem{Gaoq-lum}
M.~Ablikim {\it et al.}, BESIII Collaboration, Chin.\ Phys.\ C 39 (2015) 093001.
\end{thebibliography}
\end{document}